\documentclass[
reprint,
amsmath,amssymb,
aps,
prb,
]{revtex4-2}

\usepackage{graphicx}
\usepackage{bm}
\usepackage{color}
\usepackage{ulem}

\IfFileExists{portland.sty}{
  \usepackage{portland}
}{}

\begin{document}

\title{
Reentrant superconductivity from competing spin-triplet instabilities
}

\author{Jun Goryo}
\email{jungoryo@hirosaki-u.ac.jp}
\affiliation{
Department of Mathematics and Physics, Faculty of Science and Technology,
Hirosaki University, Hirosaki, 036-8561, Japan
}

\date{\today}

\begin{abstract}
Reentrant superconductivity in strong magnetic fields challenges the
conventional expectation that magnetic fields necessarily suppress
superconductivity.
We show that reentrant 
superconducting instability can arise from the competition
between spin-unpolarized and spin-polarized superconducting 
channels.
Using a minimal Ginzburg--Landau theory with two coupled spin-triplet order
parameters, we demonstrate that a magnetic field can reorganize the hierarchy
of superconducting instabilities, yielding a characteristic reentrant
instability curve over a broad parameter range. 
\end{abstract}

\maketitle

\section{Introduction}

Superconductivity is commonly regarded as fragile against magnetic fields due to
orbital depairing and Pauli paramagnetic effects~\cite{de2018superconductivity}. 
Orbital depairing generally affects both spin-singlet and spin-triplet
superconductors. In contrast, the Pauli paramagnetic effect universally suppresses
opposite-spin pairing, while its influence on spin-triplet states
depends sensitively on the spin structure of the order parameter.
As a result, spin-triplet superconductors can exhibit qualitatively
different magnetic-field responses~\cite{sigrist1991phenomenological}.

In spin-triplet systems, the internal spin structure of the order parameter permits
distinct superconducting instabilities that differ in their spin polarization,
whose relative stability generally depends on the direction of the applied magnetic
field. Depending on the relative phase and amplitude of the triplet components, the
dominant superconducting instability may be either spin-unpolarized
or spin-polarized in the sense that the condensate acquires a finite value of the
spin-polarization parameter $\Xi$ (see Eq.~(\ref{Xi})), leading to fundamentally
different couplings to external magnetic fields.
As a consequence, a magnetic field can do more than simply suppress
superconductivity: it may instead reorganize the hierarchy of competing
superconducting instabilities depending on its crystallographic orientation.
Such a reorganization naturally opens the possibility of nonmonotonic
behavior of the superconducting instability under magnetic fields,
including reentrant behavior.

Reentrant superconductivity, where superconductivity is suppressed at intermediate fields 
but re-emerges at higher fields, has long been discussed in several distinct physical contexts.
Early theoretical work considered exchange-field compensation in magnetic
materials (the Jaccarino--Peter mechanism)~\cite{Jaccarino1962},
while field-induced superconductivity in high magnetic fields has also been
studied in low-dimensional conductors~\cite{Lebed1986}.
More recently, magnetic-field-enhanced superconductivity has been discussed
in ferromagnetic and uranium-based superconductors, where internal magnetism
can reorganize the superconducting state~\cite{Mineev2017}.

Experimentally, reentrant superconductivity has been observed in several
uranium-based compounds that are considered candidates for spin-triplet
superconductivity~\cite{doi:10.7566/JPSJ.88.022001,Ran2019,LewinReview2024,Wu2024,Frank2024}.
Recent studies of UTe$_2$ have revealed increasingly rich high-field
superconducting phenomena, including a high-field superconducting 
halo~\cite{LewinScience2025}, and connections between high-field and
high-pressure superconducting phases~\cite{Vasina2025}.
These observations highlight that magnetic fields can play a constructive role
in stabilizing superconductivity rather than acting solely as a pair-breaking
perturbation.

Most theoretical explanations of reentrant superconductivity focus on specific microscopic ingredients, such as
particular band structures, magnetic interactions, or dimensional
effects~\cite{Jaccarino1962,Lebed1986,Agterberg1998,SHAHZADI201818, Mineev2017, mineev2020reentrant, PauliUnlimited2026}.  
In this work, we demonstrate that reentrant superconducting
instabilities can arise naturally from the competition between triplet
superconducting channels with different spin structures.
The resulting reentrant behavior depends quantitatively on effective
parameters such as the internal Josephson coupling, magnetic coupling,
orbital pair breaking, and the bare transition-temperature splitting
between the competing order-parameter components, but does not require
fine tuning to a narrow parameter regime.

\section{Ginzburg--Landau Theory}

Our analysis is based on a minimal Ginzburg--Landau description with two coupled triplet order parameters.  
We consider two complex order parameters, $(\Delta_1,\Delta_2)$, that describe coupled spin-triplet pairing channels.
The condensate spin polarization is characterized by 
\begin{equation}
i \Xi = (\Delta_1 \Delta_2^* - \Delta_2 \Delta_1^*),
\label{Xi}
\end{equation}
where $\Xi=0$ corresponds to a spin-unpolarized superconducting 
instability, while $\Xi\neq0$ characterizes a spin-polarized 
instability~\cite{sigrist1991phenomenological}. 
The quantity $\Xi$ thus directly controls the magnetic-field response 
at the level of linearized superconducting instabilities. 

The minimal linearized Ginzburg--Landau free-energy density reads
\begin{align}
f &=
\alpha_1 |\Delta_1|^2 + \alpha_2 |\Delta_2|^2
+ \varepsilon \left( \Delta_1^* \Delta_2 + \Delta_2^* \Delta_1 \right)
\nonumber\\
&\quad
- \gamma H \Xi
+ \delta H^2 (|\Delta_1|^2 + |\Delta_2|^2)
+ K \sum_{a=1,2} |{\bf D}\Delta_a|^2. 
\end{align}
The phenomenological parameters appearing in the Ginzburg--Landau
functional have simple physical interpretations.
The coefficients $\alpha_{1,2}$ control the proximity to the
superconducting instability in the two pairing channels.
Within the standard Ginzburg--Landau framework they vary linearly with
temperature, $\alpha_i = a_i (T-T_{ci})$ ($i=1,2$), where $T_{c1}$ and
$T_{c2}$ denote the bare transition temperatures of the two channels.
The quadratic mixing term $\varepsilon$ represents an internal
Josephson coupling between two non-orthogonal triplet basis
functions belonging to the same irreducible representation.
Microscopically, such a coupling arises from pair-scattering
processes that transfer Cooper pairs between the two pairing
channels. When the corresponding basis functions are not orthogonal,
pair scattering generically mixes the two components and
produces a finite $\varepsilon$, in close analogy with the
interband Josephson coupling in multiband superconductors.
The coefficient $\gamma$ describes the linear coupling between
the magnetic field and the spin polarization of the condensate,
while $\delta$ accounts for quadratic magnetic-field suppression
of superconductivity.
The coefficient of the gradient term $K$ determines the orbital response
to the magnetic field through the covariant derivative.
In a uniform magnetic field this term leads to Landau-level
quantization, and near the upper critical field the instability
is determined by the lowest Landau level.

The quadratic mixing term $\varepsilon$ favors fixing the relative phase between 
$\Delta_1$ and $\Delta_2$ to $0$ or $\pi$ and stabilizes a spin-unpolarized instability at low magnetic fields.
In contrast, the magnetic-field coupling $\gamma$ favors fixing the relative phase to $\pi/2$ or $-\pi/2$ and supports a spin-polarized instability at high fields.
This incompatibility leads to a phase frustration, which we will see is the essence of the reentrant behavior of $H_{c2}$.
Before addressing this point, however, we demonstrate in the next section that 
the coexistence of $\varepsilon$ and $\gamma$ is generically allowed.

The purpose of the present model is not to provide a complete
Ginzburg--Landau description of a specific material.
Rather, it isolates the minimal ingredients required for reentrant
superconducting instabilities, namely the coexistence of the internal
Josephson coupling $\varepsilon$ and the field-induced spin
polarization term $\gamma$.
Additional effects such as orbital anisotropy or band-dependent
gradient terms may modify the 
the results quantitatively, but are not essential for the existence of 
the mechanism discussed here.

\section{Coexistence of $\varepsilon$ and $\gamma$}
\label{coexistence}
 
The commonly discussed case of a two-dimensional irreducible representation
corresponds to the special limit $\alpha_1=\alpha_2$ and $\varepsilon=0$, in which
the two order-parameter components are symmetry-degenerate and mutually
orthogonal.
In contrast, a finite $\varepsilon$ requires two non-orthogonal spin-triplet
basis functions belonging to the same one-dimensional irreducible
representation, a situation that can arise due to spin--orbit coupling, reduced
crystal symmetry, or a multiband structure.

As discussed in connection with Eq.~(1), the coefficient $\gamma$ describes a linear coupling between the magnetic field and the spin polarization of the superconducting condensate.
Accordingly, $\gamma$ is allowed whenever multiple spin-triplet components with different spin character coexist and can be polarized by the field.
In the following, we therefore treat $\gamma$ as an effective phenomenological parameter, assuming that its presence is dictated by the spin structure of the condensate.

As a concrete example, we take
\begin{equation}
\bm d_1(\bm k)=\hat{\bm a}\,k_b,
\qquad
\bm d_2(\bm k)=\hat{\bm c}\,k_b ,
\end{equation}
where \(\hat{\bm a}\) and \(\hat{\bm c}\) are unit vectors in spin
space, and \(k_b\) denotes the momentum component along the \(b\) axis.
Under the reduced symmetry, both $\bm d_1$ and $\bm d_2$ may belong to the same one-dimensional irreducible representation.
In particular, this situation is intended to model the case of UTe$_2$ under a magnetic field parallel to 
$\bm b$ axes, where the crystal symmetry is lowered to $C_{2h}$ and both $\bm d_1$ and $\bm d_2$ are assumed to belong to the $B_u$ representation, forming two non-orthogonal basis functions within 
the same one-dimensional irreducible representation~\cite{Ran2019,Wu2024,Frank2024,LewinReview2024}.
In this situation, the quadratic mixing term $\varepsilon$ is naturally allowed and generically nonzero.

The field-induced coupling $\gamma$ is proportional to a Fermi-surface average of the spin polarization associated with the condensate, which involves the quantity
$\langle \bm d_1 \times \bm d_2 \rangle_{\mathrm{FS}}$.
For the present choice of basis functions, this average evaluates to
\begin{equation}
\langle \bm d_1 \times \bm d_2 \rangle_{\mathrm{FS}}
= (\hat{\bm a} \times \hat{\bm c})\,\langle k_b^2 \rangle_{\mathrm{FS}},
\label{gamma-estimate}
\end{equation}
which is generically nonzero as long as the spin vectors $\hat{\bm a}$ and $\hat{\bm c}$ are not parallel and the Fermi surface is not singular.

This example demonstrates that, within a single one-dimensional irreducible representation, 
an appropriate pair of non-orthogonal spin-triplet basis functions can yield a finite condensate spin polarization and hence allow a nonzero $\gamma$ term.

\section{Competing Instabilities}

To determine the superconducting instabilities, we linearize the Ginzburg--Landau equations 
near the normal state.
Linearized Ginzburg--Landau analyses of multicomponent order parameters near the critical field 
have been extensively studied~\cite{lukyanchuk1995magneticpropertiesunconventionalsuperconductors}.
The gradient terms lead to Landau-level quantization in a uniform magnetic field.
Projecting onto the lowest Landau level, the linearized equations 
reduce to a $2\times2$ eigenvalue problem,
\begin{equation}
\mathcal{M}(T,H)
\begin{pmatrix}
\Delta_1 \\
\Delta_2
\end{pmatrix}
=0,
\end{equation}
with
\begin{equation}
\mathcal{M}(T,H)=
\begin{pmatrix}
\alpha_1 + c_{\mathrm{orb}} H + \delta H^2 & \varepsilon - i\gamma H \\
\varepsilon + i\gamma H & \alpha_2 + c_{\mathrm{orb}} H + \delta H^2
\end{pmatrix}, 
\end{equation}
and the term $ c_{\mathrm{orb}} H$ denotes the cyclotron energy. 
The eigenvalues are
\begin{equation}
\lambda_{\pm}(T,H)
=
\bar{\alpha}
+ c_{\mathrm{orb}} H
+ \delta H^2
\pm
\sqrt{
\left(\frac{\Delta\alpha}{2}\right)^2
+ \varepsilon^2
+ \gamma^2 H^2
}, 
\label{eigenvalues}
\end{equation}
where $\bar{\alpha}=(\alpha_1+\alpha_2)/2$ and $\Delta\alpha=\alpha_1-\alpha_2$.  
A superconducting instability emerges when the lower eigenvalue $\lambda_-(T,H)$ vanishes.

\begin{figure}
\includegraphics[width=\linewidth]{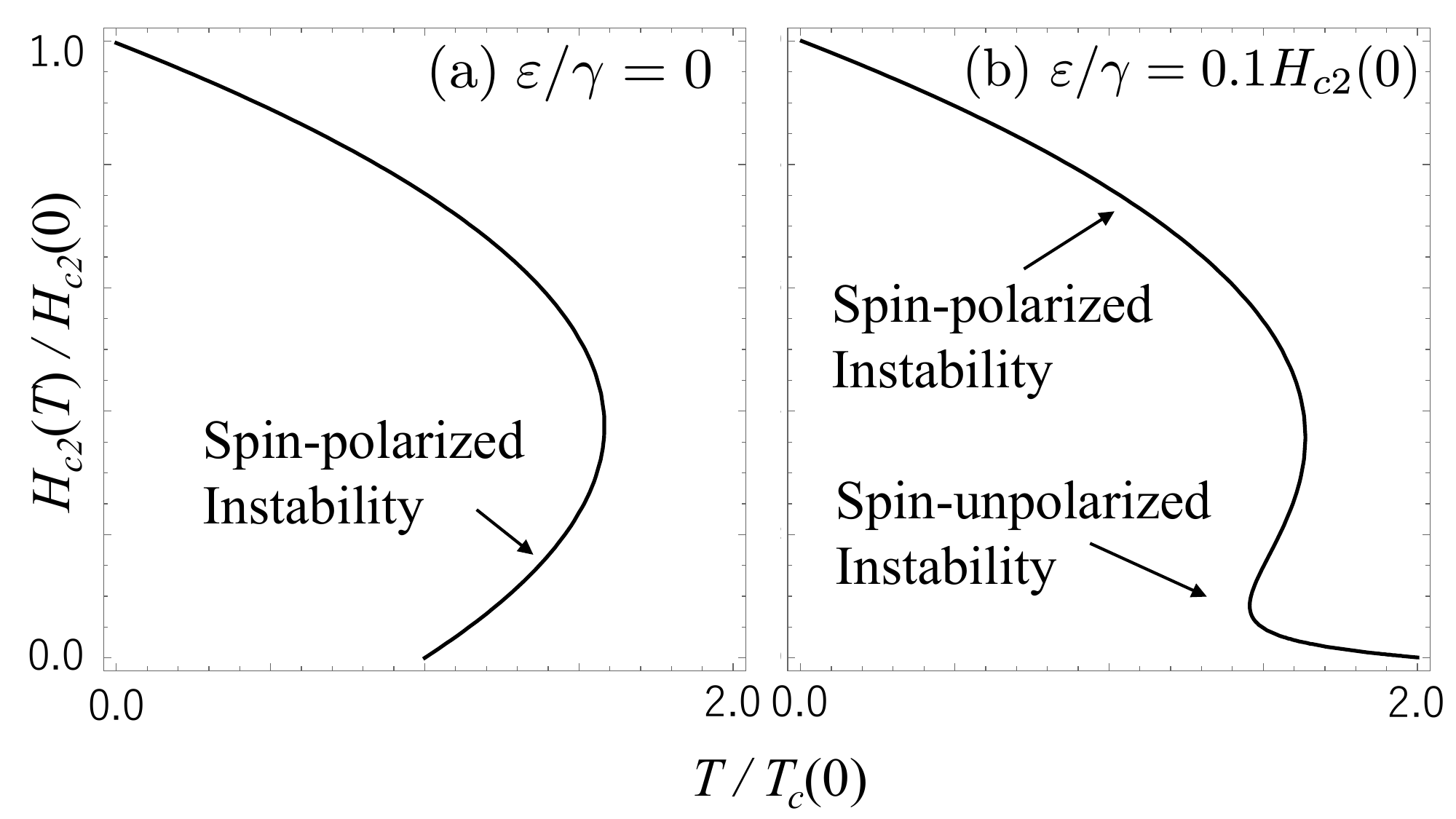}
\caption{
Representative critical-field curves $H_{c2}(T)$ obtained from the linear
instability condition $\lambda_-(T,H)=0$ [Eq.~(\ref{eigenvalues})]. 
Here, $T_c(0)$ and $H_{c2}(0)$ denote the zero-field transition temperature
and the zero-temperature upper critical field, respectively, evaluated at
$\varepsilon = \Delta \alpha=0$, and $T_{c1}=T_{c2}$.
The solid line represents the superconducting instability line, along which 
the dominant spin structure of the instability changes.
(a) $\varepsilon/\gamma =0$ (a symmetry-degenerate limit), where the instability is rapidly
reorganized toward the spin-polarized channel.
(b) $\varepsilon/\gamma=0.1 H_{c2}(0)$, where the internal Josephson coupling 
enhances the spin-unpolarized
instability at low fields, producing a reentrant structure.
For the numerical illustration we use representative parameter values
($\Delta\alpha=0$, $\gamma/c_{\mathrm{orb}}=1.2$, $\delta=0.1 c_{\rm orb}/H_{c2}(0)$),
while $\varepsilon/\gamma$ is varied as indicated in the panels.
Because our analysis is restricted to linearized superconducting instabilities,
it does not determine a phase boundary between the spin-unpolarized and spin-polarized
states. See also FIG.~\ref{normalized-Xi}.}
\label{reentrant-curve}
\end{figure}

\begin{figure}
\includegraphics[width=\linewidth]{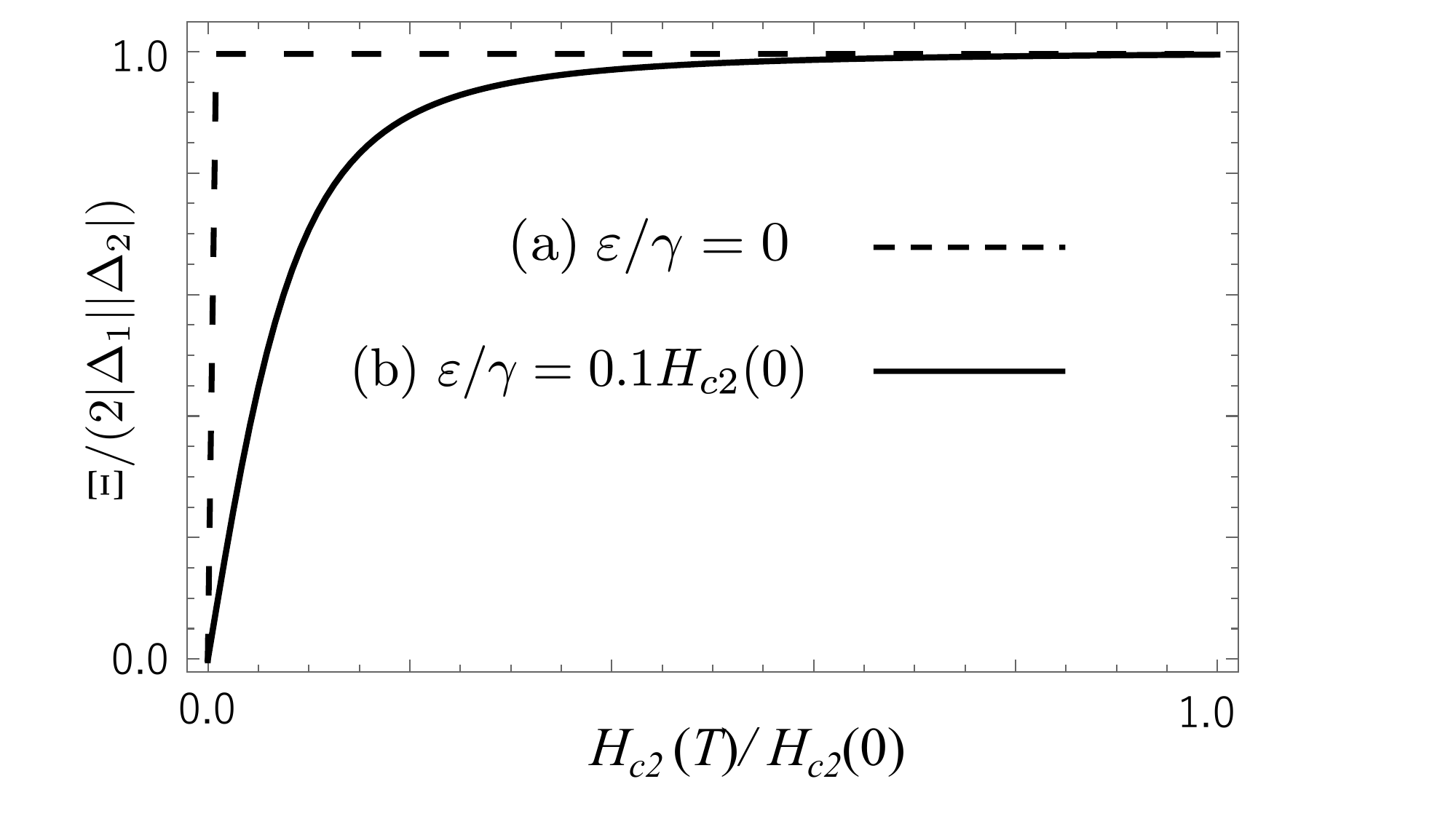}
\caption{Variation of $\Xi/(2 |\Delta_1||\Delta_2|)=\sin \phi_{12}$ on the critical field 
curves $H_{c2}(T)$. The graph shows that the magnetic field immediately favors 
the spin-polarized instability when (a) $\varepsilon/\gamma=0$, whereas 
spin-unpolarized instability persists when (b) $\varepsilon/\gamma=0.1 H_{c2}(0)$. 
This persistence caused by $\varepsilon$ is crucial for the reentrant behavior. }
\label{normalized-Xi}
\end{figure}

FIG.~\ref{reentrant-curve} shows representative solutions of the linear instability condition 
$\lambda_-(T, H)=0$, corresponding to a single instability line along which 
the dominant instability switches between different spin structures.
In the symmetry-degenerate limit $\varepsilon=0$, the magnetic field immediately favors the spin-polarized instability \textbf{[}$\phi_{12}\simeq \pm \pi/2$\textbf{]}, where $\phi_{12}=\tan^{-1}(\gamma H/ \varepsilon)$ is the relative phase between $\Delta_1$ and $\Delta_2$, yielding a convex $H_{c2}(T)$ curve.
By contrast, a finite internal Josephson coupling $\varepsilon\neq0$ stabilizes the spin-unpolarized instability 
\textbf{[}$\phi_{12}\simeq 0, \pi$\textbf{]} at low fields, thereby increasing
$T_c(H=0)$ and enabling a finite-field instability window at intermediate temperatures. 
See also FIG.~\ref{normalized-Xi}. 

\begin{figure}[h]
\includegraphics[width=\linewidth]{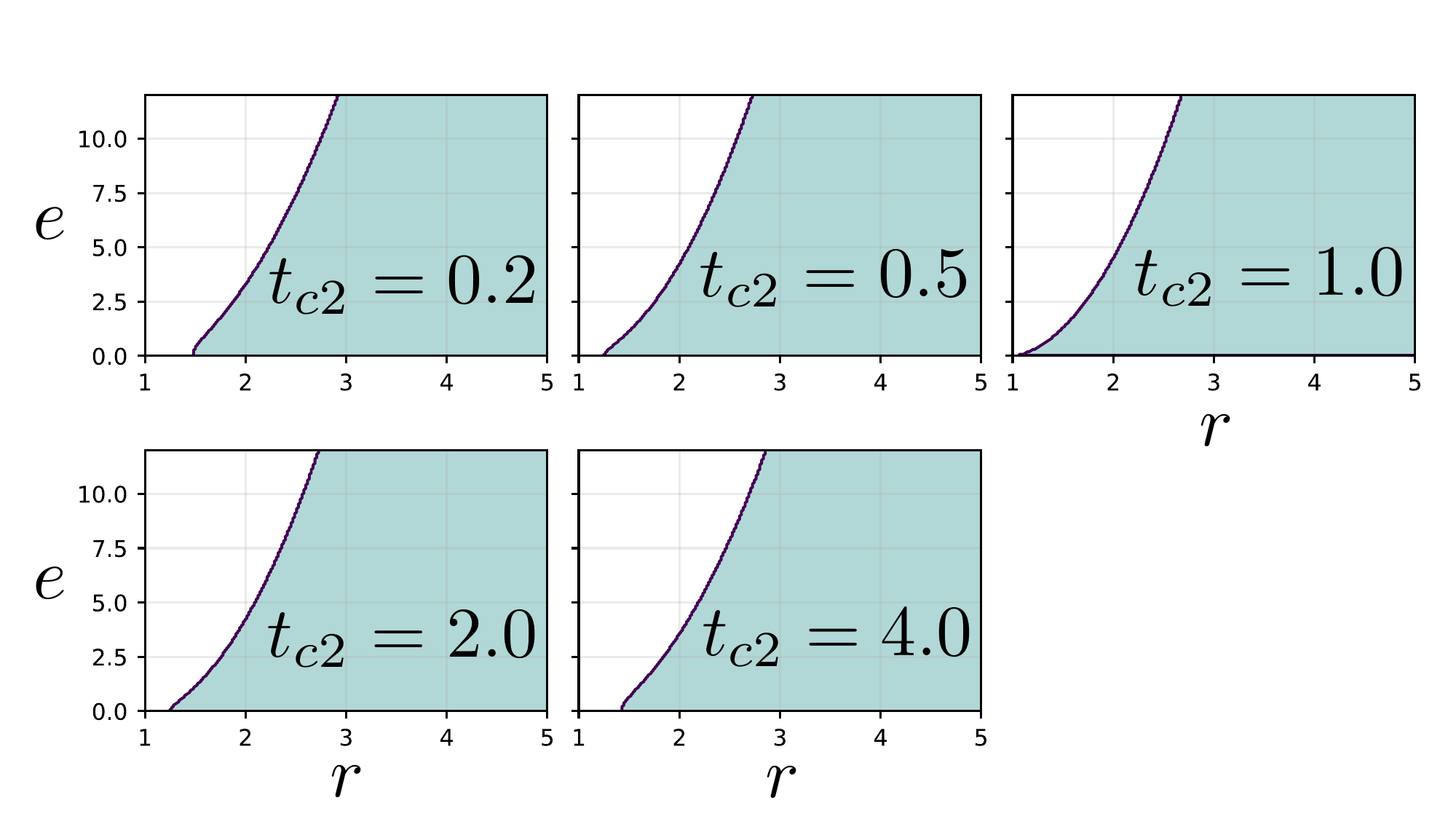}
\caption{
Reentrant region depicted by shadows in the \((r,e)\) plane for several values of
\(t_{c2}\). The parameters are fixed to \(a=1\) and \(d=0.1\) with the assumption $a_1/a_2=t_{c2}$. 
The reentrant region is largest near \(t_{c2}=1\), corresponding to the
nearly degenerate bare transition temperatures, while a finite splitting
(\(t_{c2}\neq 1\)) slightly suppresses the reentrant behavior. Detailed discission is given in Appendix \ref{app:reentrant}. 
}
\label{Reentrant_region}
\end{figure}

FIG.~\ref{Reentrant_region} shows the parameter-space maps of the
reentrant region in the \((r,e)\) plane for several values of
\(
t_{c2}=T_{c2}/T_{c1}
\),
where
\(
r=\gamma/c_{\rm orb}
\)
 and 
\(
e=\varepsilon (c_{\rm orb} H_0)^{-1}
\)
.
The parameters are fixed to
\(
a=a_2 T_{c1}(c_{\rm orb} H_0)^{-1}=1.0
\)
and
\(
d=\delta H_0^2/c_{\rm orb}=0.1
\).
For simplicity, we assume
\(
a_1 T_{c1}=a_2 T_{c2}
\),
which does not affect the essential features of the present discussion.
Here, \(H_0\) denotes the zero-temperature upper critical field in the
limit
\(
\delta=\Delta\alpha=\epsilon=\gamma=0
\).
Therefore, \(H_0\) differs slightly from 
\(H_{c2}(0)\) shown in FIG.~\ref{reentrant-curve}.
Further details of the analysis are given in
Appendix~\ref{app:reentrant}.

The maps suggest that the reentrant structure is most easily realized near \(t_{c2}=1\), where
the two bare transition temperatures are nearly degenerate.  A finite
splitting, \(t_{c2}\neq 1\), slightly shrinks the reentrant region.  Nevertheless, the
region remains rather broad in the \((r,e)\) plane, indicating that the
reentrant behavior is not a fine-tuned feature.

The key result established here is that reentrant superconducting
instabilities can already emerge at the level of competing
superconducting channels, provided that multiple spin-triplet
order-parameter components are coupled.
In this sense, the present theory identifies a minimal phenomenological
structure by which reentrant behavior can arise from instability
competition.  At the same time, the appearance and extent of the reentrant region are
controlled quantitatively by effective parameters such as the
internal Josephson coupling, magnetic coupling, orbital pair breaking, and
the bare $T_c$ splitting between different order-parameter components.

Within the present linearized theory, we determine the onset of the
superconducting instability from the normal state; determination of the
stable superconducting phase below the upper critical field requires
higher-order terms in the Ginzburg-Landau free energy.

\color{black}

\section{Paramagnetic Pair Breaking}

At the phenomenological level, the paramagnetic effect is represented
by the positive \(H^2\) contribution proportional to \(\delta\) in the
quadratic part of the free energy.
Such a term suppresses superconducting instabilities irrespective of
their spin structure.

In contrast, the Zeeman coupling to the condensate spin polarization,
proportional to \(\gamma\), acts selectively on spin-polarized
superconducting channels and progressively stabilizes them relative to
the spin-unpolarized branch as the magnetic field increases.
When \(\gamma\) exceeds the orbital pair-breaking scale
\(c_{\rm orb}\), this stabilization further leads to field-enhanced
superconductivity, as shown in FIG.~\ref{reentrant-curve}(a).
The competition between the orbital/paramagnetic suppression and the
field-induced stabilization reorganizes the hierarchy of superconducting
instabilities as a function of magnetic field.

Furthermore, in the presence of the internal Josephson coupling
\(\varepsilon\), this field-induced reorganization can evolve into a
reentrant superconducting instability, as shown in FIG.~\ref{reentrant-curve}(b).

\section{Summary and Discussions}

In summary, we have identified a minimal structure 
that can generate reentrant superconducting instabilities. 
The mechanism originates from a frustration of the relative phase between
order-parameter components caused by the competition between
the internal Josephson coupling $\varepsilon$ and
the Zeeman coupling $\gamma$.
In this sense, when reentrant superconducting behavior is observed in a
system where both \(\varepsilon\) and \(\gamma\) coexist, the mechanism
proposed here provides a natural phenomenological interpretation.

Recent reports of magnetic-field-enhanced and reentrant superconductivity in
UTe$_2$~\cite{Ran2019,LewinReview2024,Wu2024,Frank2024,LewinScience2025,Vasina2025}, 
as well as in uranium-based ferromagnetic
superconductors such as UGe$_2$, URhGe, and UCoGe~\cite{doi:10.7566/JPSJ.88.022001},
provide a natural experimental motivation for the present work. 
As mentioned in Section \ref{coexistence}, UTe$_2$ may therefore be a plausible candidate for the present mechanism.
A comprehensive overview of these uranium-based superconductors can be found in
recent reviews~\cite{Mineev2017,LewinReview2024}.

Theories for reentrant superconductivity have long been discussed in several
distinct physical contexts.
Early work considered exchange-field compensation in magnetic
materials (the Jaccarino--Peter mechanism)~\cite{Jaccarino1962}.
Field-induced superconductivity at high magnetic fields has also been discussed
in low-dimensional conductors~\cite{Lebed1986}.
In addition, magnetic-field-induced reorganization of multicomponent
superconducting order parameters has been studied in various
unconventional superconductors~\cite{Agterberg1998,SHAHZADI201818}.
Mineev has shown that reentrant behavior in UTe$_2$ can arise from a quasi-two-dimensional
electronic structure, where orbital pair breaking is strongly suppressed at high
magnetic fields~\cite{mineev2020reentrant}. 
High-field multiband superconductivity beyond the conventional Pauli limit have 
also recently appeared~\cite{PauliUnlimited2026}.

Against the background of these ongoing discussions, the present work
suggests that multicomponent spin-triplet pairing can provide a simple
phenomenological route to reentrant behavior over a broad parameter
range. Determining the full superconducting phase diagram generally requires
higher-order terms in the Ginzburg--Landau free energy.
On the other hand, as we have shown, the emergence of a reentrant upper
critical field curve can already be captured at the level of
linearized superconducting instabilities from the normal state, without
detailed knowledge of the ordered phase.
The present work is therefore not intended as a complete description of
any specific material, but rather as a minimal phenomenological
framework.  

While recent BdG studies report multiple transition temperatures at fixed
magnetic field due to state-level multivalued solutions~\cite{li2025nonunitaryspintripletsuperconductorszeeman},
the present work addresses a different phenomenon: the emergence of multiple
upper critical fields at fixed temperature arising from the competition between
nearly degenerate triplet instabilities.

Although the present analysis has focused on spin-triplet pairing, the
underlying mechanism is not necessarily restricted to spin-triplet
superconductors.
In principle, similar physics may arise in multicomponent systems where
an appropriate pair of basis functions capable of forming a condensate
coupled to an external field coexists within the same irreducible
representation.
For example, a spin-singlet condensate formed from two non-orthogonal basis functions
belonging to the same one-dimensional irreducible representation can
acquire a finite intrinsic orbital magnetization, equivalently a
nonzero Chern number.
Whether such realizations can lead to sufficiently strong magnetic-field
responses to produce pronounced magnetic-field enhancement or reentrant
superconductivity remains an open question.

\bibliographystyle{apsrev4-2}
\bibliography{Reentrant_SC_JPCM_2nd}

\appendix

\section{Parameter-Space Map for the Reentrant Region}
\label{app:reentrant}

In this appendix, we discuss the parameter dependence of the reentrant
behavior of the upper critical field.
For simplicity, we assume
\(
a_1 T_{c1}=a_2 T_{c2}
\). 
The dimensionless form of the equation
\(
\lambda_-(T,H)=0
\)
[see Eq.~(\ref{eigenvalues})] is given by
\begin{eqnarray}
 &&
 \frac{1+t_{c2}}{2}\,a t
 -1+h+d h^2
 \nonumber\\
 &&
 \qquad
 -
 \sqrt{
 \frac{(1-t_{c2})^2 a^2}{4}\,t^2
 +e^2+r^2h^2
 }
 =0 ,
 \label{eq:implicit_t_h}
\end{eqnarray}
where
\(
t=T/T_{c1}
\),
\(
t_{c2}=T_{c2}/T_{c1}
\),
and
\(
h=H/H_0
\).
The dimensionless parameters are defined as
\begin{eqnarray}
 a=\frac{a_2 T_{c1}}{c_{\rm orb} H_0},
 \qquad
 d=\frac{\delta H_0^2}{c_{\rm orb}},
 \qquad
 \nonumber\\
 e=\frac{\epsilon}{c_{\rm orb} H_0},
 \qquad
 r=\frac{\gamma}{c_{\rm orb}} .
\end{eqnarray}

We examine the field-dependence of the transition temperature $t=t(h)$ determined 
by the condition (\ref{eq:implicit_t_h}). 
We are interested in the branch which, as \(h\) increases from zero,
first decreases, then exhibits a local minimum and a local maximum,
and finally approaches \(-\infty\).

For convenience, define
\begin{eqnarray}
 A=\frac{1+t_{c2}}{2}\,a ,
 \qquad
 B=\frac{(1-t_{c2})^2 a^2}{4},
  \qquad
 \nonumber\\
 S(h)=-1+h+d h^2 .
\end{eqnarray}
Eq. \eqref{eq:implicit_t_h} can then be written as
\begin{equation}
 A t+S(h)
 =
 \sqrt{B t^2+e^2+r^2h^2}.
 \label{eq:implicit_t_h-2}
\end{equation}

Squaring both sides yields
\begin{equation}
 (A^2-B)t^2
 +2A S(h)t
 +S(h)^2-e^2-r^2h^2
 =0 .
 \label{eq:quadratic_t}
\end{equation}
Since
\begin{equation}
 A^2-B
 =
 \frac{
 a^2
 \left[
 (1+t_{c2})^2-(1-t_{c2})^2
 \right]
 }{4}
 =
 a^2 t_{c2},
\end{equation}
Eq.~\eqref{eq:quadratic_t} becomes
\begin{equation}
 a^2 t_{c2}\, t^2
 +2A S(h)t
 +S(h)^2-e^2-r^2h^2
 =0 .
\end{equation}

Hence the physically relevant branch is
\begin{equation}
 t_+(h)
 =
 \frac{
 -A S(h)
 +
 \sqrt{
 B S(h)^2
 +a^2 t_{c2}\left(e^2+r^2h^2\right)
 }
 }{
 a^2 t_{c2}
 } .
 \label{eq:tplus_branch}
\end{equation}

For large \(h\), one finds
\begin{equation}
 t_+(h)\sim
 \frac{
 -A+\sqrt{B}
 }{
 a^2 t_{c2}
 }
 d h^2 .
\end{equation}
Since \(t_{c2}>0\), the coefficient is
negative, implying
\begin{equation}
 t_+(h)\to -\infty
 \qquad
 (h\to\infty).
\end{equation}

We next derive the condition for the existence of a local minimum and a
local maximum.  Differentiating Eq.~\eqref{eq:implicit_t_h} with respect
to \(h\), we obtain
\begin{eqnarray}
 \left(
 A
 -
 \frac{B t}{\sqrt{B t^2+e^2+r^2h^2}}
 \right)
 \frac{dt}{dh}
 +
 1+2dh
 \nonumber\\
 -
 \frac{r^2h}{\sqrt{B t^2+e^2+r^2h^2}}
 =0.
\end{eqnarray}
Hence extrema satisfy
\begin{equation}
 1+2dh
 =
 \frac{r^2h}{\sqrt{B t^2+e^2+r^2h^2}} .
 \label{eq:extremum_condition}
\end{equation}

Since
\begin{equation}
 \sqrt{B t^2+e^2+r^2h^2}>rh ,
\end{equation}
a necessary condition for the existence of extrema is
\begin{equation}
 r>1.
 \label{condition-r}
\end{equation}
Moreover, all extrema must lie within
\begin{equation}
 0<h<\frac{r-1}{2d}.
 \label{condition-roots}
\end{equation}

To eliminate \(t\), define
\begin{equation}
 W(h)=\frac{r^2h}{1+2dh}.
\end{equation}
At an extremum,
\begin{equation}
 W(h)
 =
 \sqrt{B t^2+e^2+r^2h^2},
\end{equation}
while Eq.~\eqref{eq:implicit_t_h}, equivarently Eq.~\eqref{eq:implicit_t_h-2}, gives
\begin{equation}
 A t+S(h)=W(h),
\end{equation}
hence
\begin{equation}
 t=\frac{W(h)-S(h)}{A}.
\end{equation}

Substituting this into
\(
W(h)^2=B t^2+e^2+r^2h^2
\),
we obtain a single equation for \(h\):
\begin{equation}
 \Phi(h)=0 ,
 \label{eq:Phi_condition}
\end{equation}
where
\begin{align}
 \Phi(h)
 &=
 a^2 t_{c2}\, W(h)^2
 +2B S(h)W(h)
 \nonumber\\
 &\quad
 -B S(h)^2
 -A^2\left(e^2+r^2h^2\right)
 \nonumber\\
 &=
 a^2 t_{c2}
 \left(
 \frac{r^2h}{1+2dh}
 \right)^2
 \nonumber\\
 &\quad
 +2B S(h)
 \left(
 \frac{r^2h}{1+2dh}
 \right)
 \nonumber\\
 &\quad
 -B S(h)^2
 -A^2\left(e^2+r^2h^2\right).
  \label{eq:Phi_definition}
\end{align}

Therefore the curve \(t_+(h)\) exhibits the structure
\begin{eqnarray}
 \text{decrease}
 \;\longrightarrow\;
 \text{local minimum}
 \;\longrightarrow\;
 \text{increase}
  \qquad
 \nonumber\\
 \;\longrightarrow\;
 \text{local maximum}
 \;\longrightarrow\;
 -\infty
 \nonumber
\end{eqnarray}
if and only if Eq.~\eqref{condition-r} is satisfied  
and Eq.~\eqref{eq:Phi_condition} possesses exactly two positive solutions \(h_1<h_2\) 
in the interval indicated by Eq.~\eqref{condition-roots}. 
The first root $h_1$ corresponds to the local minimum, while the second one $h_2$ 
corresponds to the local maximum.

In the case of $t_{c2}=1$, the positive \(B t^2\) contribution inside the
square root in Eq.~\eqref{eq:extremum_condition} is dropped, and we obtain the useful
necessary condition
\begin{equation}
 \max_{h>0}
 \left[
 \frac{r^2h}{\sqrt{e^2+r^2h^2}}
 -2dh
 \right]
 =r\left[ 1-\left(\frac{2de}{r^2}\right)^{2/3} \right]^{3/2}
 >1.
\end{equation}
This gives
\begin{equation}
 r>1,
 \qquad
 0<d<
 \frac{r^2}{2e}
 \left(1-r^{-2/3}\right)^{3/2}.
\end{equation}
Thus, for a given \(r\), the parameter \(e\) has a finite upper bound
for the appearance of the reentrant structure.

Fig.~\ref{Reentrant_region} shows the parameter map 
indicating the region where the reentrant conditions are satisfied. 
The reentrant structure is most easily realized near \(t_{c2}=1\), where
the two bare transition temperatures are nearly degenerate.  A finite
splitting, \(t_{c2}\neq 1\), introduces the \(t\)-dependent term in the
square root and slightly shrinks the reentrant region.  Nevertheless, the
region remains rather broad in the \((r,e)\) plane, indicating that the
reentrant behavior is not a fine-tuned feature.

\end{document}